\documentclass[12pt]{article}
\usepackage{latexsym}
\usepackage{amsmath,,calrsfs}
\usepackage{amsfonts}
\usepackage{amssymb}
\usepackage{amscd}
\usepackage{bbm}
\usepackage{fancybox}
\usepackage{cite}
\usepackage{amsmath,amsfonts,amsbsy}
\usepackage{pstricks,pst-node}
\usepackage[small,bf,hang]{caption2}
\usepackage{graphicx}
\usepackage{epsfig}
\usepackage{psfrag}
\usepackage{comment}

\usepackage{float}

\psset{unit=1.3cm,linewidth=.5pt,radius=.2}  

\usepackage{multirow}                     
\usepackage{float}                          
\usepackage{lscape}                         
\usepackage{bm}

\newcommand{\bb}{\bar\beta}

\newcommand{\beq}{\begin{equation}}
\newcommand{\eeq}{\end{equation}}
\newcommand{\bi}{\begin{itemize}}
\newcommand{\ei}{\end{itemize}}
\newcommand{\bt}{\begin{tabular}}
\newcommand{\et}{\end{tabular}}
\newcommand{\bc}{\begin{center}}
\newcommand{\ec}{\end{center}}

\newcommand{\be}{\begin{equation}}
\newcommand{\ee}{\end{equation}}
\newcommand{\bea}{\begin{eqnarray}}
\newcommand{\eea}{\end{eqnarray}}
\newcommand{\ba}{\begin{array}}
\newcommand{\ea}{\end{array}}

\def\bbox{{\,\lower0.9pt\vbox{\hrule \hbox{\vrule height 0.2 cm
\hskip 0.2 cm \vrule height 0.2 cm}\hrule}\,}}
\newcommand{\dsl}{\pa \kern-0.5em /}

\font\mybb=msbm10 at 10pt
\def\bb#1{\hbox{\mybb#1}}
\def\bZ {\bb{Z}}

\def\bL {\bb{L}}
\def\bG {\bb{G}}

\newcommand{\bfa}{\mbox{\boldmath $\alpha$}}
\newcommand{\tbfa}{\mbox{\boldmath $\tilde \alpha$}}

\makeatletter \@addtoreset{equation}{section} \makeatother

\def\slashchar#1{\setbox0=\hbox{$#1$}           
   \dimen0=\wd0                                 
   \setbox1=\hbox{/} \dimen1=\wd1               
   \ifdim\dimen0>\dimen1                        
      \rlap{\hbox to \dimen0{\hfil/\hfil}}      
      #1                                        
   \else                                        
      \rlap{\hbox to \dimen1{\hfil$#1$\hfil}}   
      /                                         
   \fi}

\begin{document}

\begin{titlepage}
\begin{center}

\hfill  UMTG-30, DAMTP-2013-28 

\vskip 1.5cm

{\Large \bf  Equivalence of  3D Spinning String and Superstring}

\vskip 1cm

{\bf Luca Mezincescu\,${}^1$, Alasdair J. Routh\,${}^2$ and 
Paul K.~Townsend\,${}^2$} \\

\vskip 25pt

{\em $^1$ \hskip -.1truecm
\em Department of Physics
University of Miami, \\ Coral Gables, FL 33124, USA\vskip 5pt }

{email: {\tt L.Mezincescu@server.physics.miami.edu}} \\

\vskip .4truecm

{\em $^2$ \hskip -.1truecm
\em  Department of Applied Mathematics and Theoretical Physics,\\ Centre for Mathematical Sciences, University of Cambridge,\\
Wilberforce Road, Cambridge, CB3 0WA, U.K.\vskip 5pt }

{email: {\tt A.J.Routh@damtp.cam.ac.uk, P.K.Townsend@damtp.cam.ac.uk}} \\

\end{center}

\vskip 0.5cm

\begin{center} {\bf ABSTRACT}\\[3ex]
\end{center}


We perform a light-cone gauge quantization of the Ramond-Ramond sector of the closed spinning string in three spacetime dimensions (3D). The spectrum is Lorentz invariant and identical to that of the 3D Green-Schwarz closed superstring with ${\cal N}=2$ space-time supersymmetry, quantized in light-cone gauge.

\end{titlepage}

\newpage
\setcounter{page}{1} 

\newpage

\section{Introduction}

There are two standard formulations of the ten-dimensional (i.e. critical dimension) superstring theory. The first to be found was the ``RNS formulation'' which was obtained, as a free string theory, by removing from the combined spectrum of the  Ramond \cite{Ramond:1971gb} and Neveu-Schwarz \cite{Neveu:1971rx} spinning strings the states which do not form multiplets of spacetime supersymmetry using the Gliozzi-Scherk-Olive (GSO) projection \cite{Gliozzi:1976qd}.  An alternative light-cone gauge action with manifest spacetime supersymmetry was then proposed by Green and Schwarz  \cite{Green:1980zg} and shown by them to be equivalent (by virtue of the triality property of the ${\rm Spin}(8)$ transverse rotation group) to the light-cone gauge-fixed RNS superstring; in this alternative ``GS  formulation''  the GSO projection is transformed into the simple requirement of Ramond-type boundary conditions on the fermionic variables (i.e. periodicity for a closed string). Green and Schwarz subsequently found the covariant form of their alternative string action \cite{Green:1983wt}, and this is the natural starting point of the GS formulation of superstring theory. 

An intriguing feature of the GS formulation is that even the classical superstring  action exists only for spacetime dimensions $D=3,4,6,10$. As for the RNS string,  quantization  of the  light-cone gauge-fixed action preserves Lorentz invariance if $D=10$ but not otherwise for $D\ge4$. Recently,  two of us have shown that the Lorentz invariance, and spacetime supersymmetry, of the superstring are also preserved  when $D=3$  although the spectrum then contains particles of {\it irrational} spin \cite{Mezincescu:2010yp}.  It is not known how to recover this result by covariant quantization of the 3D  GS superstring; one difficulty  is that the covariant  wave equation for a  particle of irrational spin  requires an infinite-component wave-function  \cite{Jackiw:1990ka,Plyushchay:1990rt}. 

Apart from this difficulty  there are other well-known difficulties in the covariant quantization of the superstring in the GS formulation, so it could be useful if there were some 
3D analog of the critical dimension equivalence with the RNS superstring. The principal new result of this paper is a proof that  the 3D GS closed superstring with ${\cal N}=2$ space-time supersymmetry is equivalent to the Ramond-Ramond sector of the 3D  closed spinning string, which we refer to here as the (3D) ``Ramond string''.  In other words, there is indeed a 3D analog of the critical dimension GSO projection: it involves projecting out the sectors involving NS boundary conditions.  
We establish this result by performing a light-cone gauge quantization of the 3D Ramond string and comparing to the analogous GS superstring results obtained previously \cite{Mezincescu:2010yp}. The spectrum is found to be identical. This means, in particular, that there are states of irrational spin in the spectrum of the 3D spinning string. 

We begin with a discussion of  the Ramond string in a Minkowski spacetime of general dimension $D$. Our procedure  contains some novelties:  we use a Hamiltonian formulation  of the action, which we first rewrite  in terms of Fourier modes and then gauge-fix to arrive at  a light-cone gauge action, which is still invariant under global reparametrizations of the worldsheet since we fix the gauge only on the transverse oscillator variables. The two constraints associated to these residual gauge invariances are the mass-shell and level-matching constraints, which become physical-state conditions in the quantum theory.  

We then focus on the $D=3$ case. All irreducible representations of the 3D Poincar\'e group are characterized by the values of the two quadratic Casimirs $-{\cal P}^2= {\cal M}^2$ and ${\cal P} \cdot {\cal J} \equiv \Lambda$,  where ${\cal P}$ is the 3-vector  translation generator and ${\cal J}$ the 3-vector Lorentz rotation generator.  In the application to string theory, 
${\cal M}^2$ and $\Lambda$ become commuting operators acting on the one-string Hilbert space, and their joint eigenvalues determine the spectrum of masses and spins. Spin is not actually defined for zero mass although there is still a distinction between bosons and fermions  \cite{Binegar:1981gv,Deser:1991mw}, and ``infinite spin'' \cite{Edgren:2005gq}. For positive mass eigenvalue $M$ we define ``spin'' to be the absolute value of the ``relativistic helicity'' $\Lambda/M$.  We establish equivalence of the 3D Ramond string to the 3D GS superstring  by showing that the light-cone gauge expressions for the operators ${\cal M}^2$ and $\Lambda$ coincide for a particular  map between the canonical variables of the two light-cone gauge string actions. 

A corollary of this result is that the worldsheet supersymmetric 3D Ramond string actually has 3D space-time supersymmetry. We use the equivalence to the 3D Green-Schwarz superstring to construct the space-time supersymmetry charges of the 3D Ramond string, in light-cone gauge. 

We should stress that the aim of this paper is to establish that two ostensibly distinct actions describing a free string in three-dimensional Minkowski space actually have 
identical physical spectra, which are Lorentz invariant because the usual Lorentz anomaly in a subcritical dimension $D$ is absent for $D=3$. We do not aim here to 
overcome any of the obstacles facing the construction of a consistent 3D string theory. In particular, we do not address the issue of interactions. However, even for a free closed string there is the issue of modular invariance of the one-loop partition function. We comment on this in our concluding section because modular invariance of  the critical dimension RNS string requires a sum over worldsheet spin structures, which requires inclusion of sectors with NS boundary conditions, whereas we consider here
the closed 3D string with only Ramond boundary conditions.

\section{Ramond string}

We start from the Hamiltonian form of the action for the closed spinning string in a $D$-dimensional Minkowski space-time; an introduction to the Hamiltonian formalism applied to the RNS string can be found in the  lectures of Henneaux in \cite{Brink:1988nh}. The canonical variables are the commuting 
$D$-vectors $(X,P)$ and the anti-commuting $D$-vectors $(\psi,\tilde\psi)$. The action is\footnote{Elimination of $P_m$ leads to the standard action for the spinning string as an  action for  worldsheet  supergravity coupled to $D$ worldsheet  ``matter''  supermultiplets \cite{Brink:1976sc,Deser:1976rb}, where $(\psi^m,\tilde\psi^m)$ are the components of the world sheet spinor  partners to $X^m$; the Lagrange multipliers become the independent components of the conformal world sheet graviton supermultiplet.}
\be\label{Ramondact}
S= \int \!dt \oint d\sigma \left\{ \dot X^m P_m + \frac{i}{2} \psi\cdot\dot\psi +\frac{i}{2}\tilde\psi\cdot\dot{\tilde\psi} - \lambda {\cal H} - \tilde\lambda \tilde {\cal H} -i\chi {\cal Q} -i\tilde\chi\tilde{\cal Q} \right\}\, ,
\ee
where the commuting Lagrange multipliers $(\lambda,\tilde\lambda)$ impose the bosonic constraints, with constraint functions
\be
{\cal H} = \frac{1}{4T} \left(P-TX'\right)^2 - \frac{i}{2} \psi\cdot \psi'  \, , \qquad \tilde{\cal H} = \frac{1}{4T} \left(P+TX'\right)^2 +\frac{i}{2}\tilde\psi\cdot\tilde\psi'\, ,  
\ee
and the anticommuting Lagrange multipliers $(\chi,\tilde\chi)$ impose the fermionic constraints, with constraint functions 
\be
{\cal Q}= \frac{1}{2\sqrt{T}} \left(P-TX'\right)\cdot \psi \, , \qquad \tilde{\cal Q}= \frac{1}{2\sqrt{T}} \left(P+TX'\right)\cdot \tilde\psi\, . 
\ee
As usual, the primes here  indicate a derivative with respect to the string coordinate $\sigma$.

We can read off from the action (\ref{Ramondact}) the non-zero Poisson Brackets (PBs) of the canonical variables; in particular, 
\be
\left\{\psi^m(\sigma),\psi^n (\sigma') \right\}_{PB} = \left\{\tilde\psi^m(\sigma),\tilde\psi^n (\sigma') \right\}_{PB} = -i\delta(\sigma-\sigma')\eta^{mn}\, , 
\ee
where $\eta$ is the Minkowski space-time metric (with ``mostly plus'' signature). 
A calculation now shows that the non-zero PBs of the untilded constraint functions are
\begin{eqnarray}
\left\{ {\cal Q}(\sigma),{\cal Q}(\sigma')\right\}_{PB} &=& -i {\cal H}(\sigma) \delta(\sigma-\sigma')\nonumber \\
\left\{ {\cal Q}(\sigma),{\cal H}(\sigma')\right\}_{PB} &=& -\left[\frac{1}{2}{\cal Q}(\sigma) + {\cal Q}(\sigma')\right] \delta'(\sigma-\sigma')\nonumber \\
\left\{{\cal H} (\sigma),{\cal H}(\sigma')\right\}_{PB} &=& \left[{\cal H}(\sigma) + {\cal H}(\sigma')\right]\delta'(\sigma-\sigma')\, . 
\end{eqnarray}
The same expressions hold for tilded quantities and there are no non-zero PBs between tilded and untilded constraint functions. From this we see that the constraints are all first-class, so the constraint functions generate gauge invariances
of the canonical variables. The non-zero infinitesimal gauge transformations are 
\begin{eqnarray}\label{gt1}
\delta X &=& \frac{1}{2T} \xi\left(P-TX'\right) + \frac{i}{2\sqrt{T}} \epsilon\psi\, , \qquad \delta P= -T\left(\delta X\right)' \nonumber \\
\delta\psi &=&  -\xi\psi' - \frac{1}{2} \xi' \psi - \frac{1}{2\sqrt{T}}\left(P-TX'\right)\epsilon\, , 
\end{eqnarray}
and 
\begin{eqnarray} \label{gt2}
\delta X &=&  \frac{1}{2T} \tilde\xi\left(P+TX'\right) + \frac{i}{2\sqrt{T}} \tilde\epsilon\tilde\psi\, , \qquad \delta P= T\left(\delta X\right)' \nonumber \\
\delta\tilde\psi &=&  \tilde\xi\tilde\psi' +\frac{1}{2} \tilde\xi' \tilde\psi - \frac{1}{2\sqrt{T}}\left(P+TX'\right)\tilde\epsilon\, , 
\end{eqnarray}
where $(\xi,\tilde\xi)$ and $(\epsilon,\tilde\epsilon)$ are, respectively, the commuting and anti-commuting parameters. 
Using these results, a  computation shows that
\begin{eqnarray}
\delta {\cal H} &=& - \left[\xi{\cal H}\right]' - \xi'{\cal H}  -\frac{i}{2} \left(\epsilon{\cal Q}\right)' -i\epsilon' {\cal Q}\nonumber \\
\delta {\cal Q} &=& -\left[\xi{\cal Q}\right]' - \frac{1}{2} \xi'{\cal Q}  - {\cal H} \epsilon\, , 
\end{eqnarray}
and
\begin{eqnarray}
\delta \tilde{\cal H} &=& \left[{\tilde \xi}\tilde{\cal H}\right]^\prime+ {\tilde \xi}^\prime \tilde{\cal H} +  \frac{i}{2}\left({\tilde \epsilon}\tilde{\cal Q}\right)^\prime+ i{\tilde \epsilon}^\prime \tilde{\cal Q}\ ,  \nonumber \\
\delta\tilde{\cal Q}&=& \left[{\tilde \xi}\tilde{\cal Q}\right]^\prime + \frac{1}{2}{\tilde \xi}^\prime \tilde{\cal Q} - \tilde{\cal H} {\tilde \epsilon}\, .
\end{eqnarray}
The action is then found to be invariant if the Lagrange multipliers transform as 
\begin{eqnarray}
\delta  \lambda &=& \dot\xi - \xi \lambda'  + \lambda\xi'  +  i\chi\epsilon \nonumber \\
\delta \chi &=& -\xi \chi' + \frac{1}{2} \xi' \chi + \dot\epsilon + \lambda \epsilon' - \frac{1}{2}\lambda'\epsilon\, , 
\end{eqnarray}
and
\begin{eqnarray}
\delta\tilde \lambda &=&{\dot {\tilde \xi}} +  {\tilde \xi}{\tilde \lambda}^{\,\prime}  -{\tilde \xi}^{\,\prime} {\tilde \lambda}+  i{\tilde \chi} {\tilde \epsilon},  \nonumber \\
\delta{\tilde \chi} &=& {\tilde \xi}{\tilde \chi}^{\,\prime}- \frac{1}{2}{\tilde \xi}^{\,\prime} {\tilde \chi } + {\dot {\tilde \epsilon}} + 
\tilde \lambda{\tilde \epsilon}^{\,\prime} - \frac{1}{2}{\tilde \lambda}^\prime {\tilde \epsilon}\, . 
\end{eqnarray}

The action (\ref{Ramondact}) is also manifestly Poincar\'e invariant, with corresponding Noether charges
\be
{\cal P}^m = \oint d\sigma P^m\, , \qquad 
{\cal J}^{mn} =  \oint \! d\sigma \left( X^m P^n - X^n P^m -i \psi^m\psi^n -i\tilde\psi^m\tilde\psi^n \right)\, . 
\ee

\subsection{Fourier space action}

It is convenient to pass to the Fourier space form of the action prior to gauge fixing. We 
need only the Ramond-Ramond sector for which the world sheet fermions are periodic; 
we call this the closed Ramond string. We shall take the parameter length of the string to be $2\pi$, so that all  worldsheet fields are periodic 
in $\sigma$ with period $2\pi$.  The appropriate Fourier series expansions of the canonical variables are 
\begin{eqnarray}\label{fourier}
P -T X' &=& \sqrt{\frac{T}{\pi}}\sum_{k\in \bZ} e^{i k \sigma} \alpha_k  \, , \qquad \psi= \frac{1}{\sqrt{2\pi}} \sum_{k\in \bZ}  e^{ ik\sigma}\,  d_k(t)\, , \nonumber \\
P +T  X'  &=& \sqrt{\frac{T}{\pi}} \sum_{k\in\bZ} e^{-i k \sigma} \tilde\alpha_k \, , \qquad \tilde\psi = \frac{1}{\sqrt{2\pi}} \sum_{k\in \bZ}  e^{-i k\sigma}\,  \tilde d_k(t)\, . 
\end{eqnarray}
It follows that 
\be
p\equiv \oint\! d\sigma P = \sqrt{4\pi T} \, \alpha_0 = \sqrt{4\pi T} \, \tilde\alpha_0\, .
\ee
The $D$-vector  variable $p(t)$  is canonically conjugate to the integration constant $x(t)$ that appears on integrating the  Fourier series for $X'$.  
We may similarly express the Lagrange multipliers as Fourier  series:
\begin{eqnarray}\label{lambdachiexp}
\lambda &=& \sum_{n \in {\bb{Z}}}e^{in\sigma}\lambda_n\, , \qquad  {\tilde\lambda }= \sum_{n \in {\bb{Z}}}e^{- in\sigma}{\tilde \lambda}_n\, , \nonumber \\
 \chi &=& \sum_{n \in \bZ}e^{in\sigma}\chi_n\, , \qquad  {\tilde \chi} = \sum_{n \in \bZ}e^{- in\sigma}{\tilde \chi}_n\, . 
\end{eqnarray}
The constraint functions now have the following Fourier series expansions:
\begin{eqnarray}
{\cal H} &=& \frac{1}{2\pi}\sum_{n \in \bZ}e^{in\sigma} L_n \, , \qquad\ 
L_n= \frac{1}{2}\sum_{k \in \bZ } \left[ \alpha_k \cdot\alpha_{n-k} -k\, d_k\cdot d_{n-k}\right], \nonumber \\
\tilde{\cal H} &=& \frac{1}{2\pi}\sum_{n \in \bZ}e^{-in\sigma}\tilde L_n \, , \qquad
{\tilde L}_n= \frac{1}{2}\sum_{k \in \bZ}\left[\tilde\alpha_k \cdot\tilde\alpha_{n-k} -k\, {\tilde d}_k\cdot {\tilde d}_{n-k}\right]\, , 
\end{eqnarray}
and 
\begin{eqnarray}
{\cal Q} &=& \sqrt{\frac{1}{8\pi^2}}\sum_{n \in \bZ} e^{in\sigma}G_n \, , \qquad 
G_n = \sum_{k \in \bZ} \alpha_k\cdot d_{n-k}\, , \nonumber \\
\tilde{\cal Q} &=& \sqrt{\frac{1}{8\pi^2}}\sum_{n \in \bZ} e^{- in\sigma}{\tilde G}_n \, , \qquad
{\tilde G}_n = \sum_{k \in {\bb{Z}} }{\tilde \alpha}_k\cdot {\tilde d}_{n-k} \, .
\end{eqnarray}

Using these Fourier series expressions, we find that the Fourier space action is
\begin{eqnarray}
S &=& \int \! dt \left\{ \dot x^m p_m + \frac{i}{2} \left(d_0\cdot \dot d_0 + 
\tilde d_0\cdot \dot {\tilde d}_0 \right)
+ i \sum_{k=1}^\infty \left[ \frac{1}{k} \left(\dot\alpha_k \cdot \alpha_{-k} + 
\dot{\tilde\alpha}_k \cdot \tilde\alpha_{-k}\right) + d_{-k}\cdot \dot d_k + 
\tilde d_{-k}\cdot \dot {\tilde d}_k \right] \right. \nonumber \\
&&\left. - \sum_{n\in \bZ} \left[\lambda_{-n}L_n + \tilde\lambda_{-n} \tilde L_n + \frac{i}{\sqrt{2}} \left(\chi_{-n}G_n + \tilde\chi_{-n} \tilde G_n\right) \right]
\right\}\, . 
\end{eqnarray}
This action is of course still invariant under the gauge transformations of (\ref{gt1}) and (\ref{gt2}).  If the gauge parameters are expressed as the Fourier series
\be
\xi = \sum_{n \in \bZ}e^{ in\sigma}\xi_n\, , \quad 
{\tilde\xi }= \sum_{n \in \bZ}e^{- in\sigma}{\tilde \xi}_n\, , \qquad
\epsilon = \sum_{k \in \bZ} e^{ ik\sigma}\epsilon_k, \quad  
{\tilde \epsilon} = \sum_{k \in \bZ}e^{- ik\sigma}\tilde\epsilon_k \, , 
\ee
then we find that the (non-zero) gauge transformations become
\be
\delta x = \sqrt{\frac{1}{4\pi T}}\sum_{k \in \bZ} \left[ \xi_k\alpha_{-k} +{\tilde \xi}_k{\tilde \alpha}_{-k} + \frac{i}{\sqrt{2}}\left(\epsilon_k d_{-k} +  \tilde\epsilon_k \tilde d_{-k} \right)\right]\, , 
\ee
and
\begin{eqnarray}
\delta\alpha_n &=& -  n\sum_{k \in \bZ} \left(i \xi_k\alpha_{n-k} -  \frac{1}{\sqrt{2}} \epsilon_k d_{n-k}\right)\, , \nonumber \\
\delta d_n &=&  -\sum_{k \in {\bb{Z}}}\left[i\left( n -\frac{k}{2}\right)\xi_k d_{n-k} - \frac{1}{\sqrt{2}} 
\epsilon_{n - k}\, \alpha_{k}\right]\, , \nonumber \\
\delta\tilde\alpha_n &=& -  n\sum_{k \in \bZ} \left(i \xi_k\tilde\alpha_{n-k} -  \frac{1}{\sqrt{2}} \tilde\epsilon_k \tilde d_{n-k}\right)\, , \nonumber \\
\delta {\tilde d}_n &=&  -\sum_{k \in {\bb{Z}}} \left[ i\left( n -\frac{k}{2}\right){\tilde \xi}_k{\tilde d}_{n-k} -\frac{1}{\sqrt{2}}\tilde \epsilon_{n - k}\, {\tilde \alpha}_{k}\right]\, . 
\end{eqnarray}
These variations imply that
\begin{eqnarray}
\delta L_n &=& - \sum_{k \in \bZ}\left[ i(k+n)\xi_kL_{n-k} - \frac{1}{\sqrt{2}} (k +\frac{n}{2})\epsilon_k G_{n-k}\right]\, , \nonumber \\
\delta G_n &=& -\sum_{k \in \bZ}\left[ i(n +\frac{k}{2})\xi_kG_{n-k} + \sqrt{2}\, 
\epsilon_{n -k}L_{k} \right] \, , 
\end{eqnarray}
and 
\begin{eqnarray}
\delta {\tilde L}_n &=& - \sum_{k \in \bZ} \left[ i(k+n){\tilde \xi}_k{\tilde L}_{n-k} - \frac{1}{\sqrt{2}} 
(k +\frac{n}{2}){\tilde \epsilon}_k{\tilde G}_{n-k}\right] \, , \nonumber \\
\delta \tilde G_n &=& -\sum_{k \in \bZ}\left[ i(n +\frac{k}{2})\xi_k\tilde G_{n-k} + \sqrt{2}\, 
\epsilon_{n-k}\tilde L_{k} \right] \, . 
\end{eqnarray}
The gauge  transformations of the Lagrange multipliers are:
\begin{eqnarray}
\delta \lambda_n &=& {\dot \xi}_n + i \sum_{k \in {\bb{Z}}}\left[\left(2k-n\right)\xi_k\lambda_{n-k} -  \epsilon_{n-k}\chi_{k}\right]\, , \nonumber \\
\delta \chi_n &=& {\dot \epsilon}_n+ \frac{i}{2}\sum_{k \in \bZ} \left[\left(3k -n\right)\epsilon_k\lambda_{n - k} + \left(3k-2n\right) \xi_k\lambda_{n-k} \right] \, , 
\end{eqnarray}
and 
\begin{eqnarray}
\delta {\tilde \lambda}_n &=& {\dot {\tilde \xi}}_n + i \sum_{k \in {\bb{Z}}}\left[(2k-n){\tilde \xi}_k{\tilde \lambda}_{n-k} - {\tilde \epsilon}_{n-r}{\tilde \chi}_{r}\right]\, , \nonumber \\
\delta {\tilde \chi}_n &=& {\dot {\tilde \epsilon}}_n+ \frac{i}{2}\sum_{k \in \bZ} \left[(3k -n){\tilde \epsilon}_k{\tilde \lambda}_{n-k}+\left(3k-2n\right)
{\tilde \xi}_k{\tilde \lambda}_{n-k}\right]  \, . 
\end{eqnarray}

Finally, we may express the Poincar\'e generators in terms of the Fourier modes: 
\be
{\cal P}^m = p^m\, , \qquad  {\cal J}^{np} = x^np^p - x^p p^n   - i \left(d_0^n d_0^p + \tilde d_0^n\tilde d_0^p\right) +  S^{np} + \tilde S^{np}\, , 
\ee
where
\be
S^{np}  =  \sum _{k\ne0} \left[ \frac{i}{k}\alpha^{[n}_k \alpha^{p]}_{-k} -id^{[n}_{-k} { d}^{p]}_k\right]\, , \qquad
\tilde S^{np} = \sum _{k\ne0} \left[ \frac{i}{k}\tilde\alpha^{[n}_k \tilde\alpha^{p]}_{-k} -i  \tilde d^{[n}_{-k} { \tilde d}^{p]}_k\right]\, . 
\ee

\subsection{Light-cone gauge fixing} 

We now impose the gauge-fixing conditions
\be
\alpha_k^+ = \tilde\alpha_k^+ =0\, ,  \quad k\ne0\, ; \qquad d_k^+=\tilde d_k^+ =0 \, , \quad \forall k\, . 
\ee
Provided that $\alpha_0^+\ne0$, which is equivalent to $p_-\ne0$, these  conditions fix all gauge invariances\footnote{This is not true for $D=2$, in which case one cannot assume that 
$p_-\ne0$ \cite{Bardeen:1975gx} but once $D>2$ the issues involved in the assumption that $p_-\ne0$ are the same for any $D$; in particular, they are the same for $D=3$ as for $D=10$.} 
except those generated by $(L_0,\tilde L_0)$.  At the same time, all the other constraints may be solved for $(\alpha_n^-,\tilde\alpha_n^-)$ and $(d_n^-,{\tilde d}_n^-)$:
\begin{eqnarray}\label{minusosc}
\alpha_n^- &=& - \frac{1}{\alpha_0^+} \bL_n \quad (n\ne0)\, ,  \qquad d_n^- = - \frac{1}{\alpha_0^+} \bG_n \quad  (\forall n)\, , \nonumber \\
\tilde\alpha_n^- &=& - \frac{1}{\alpha_0^+} \tilde\bL_n \quad (n\ne0)\, ,  \qquad \tilde d_n^- = - \frac{1}{\alpha_0^+} \tilde\bG_n \quad  (\forall n)\, , 
\end{eqnarray}
where 
\be\label{transverseL}
\bL_n = \frac{1}{2}\sum_{k\in\bZ} \left[ \bfa_k\cdot \bfa_{n-k} + k\, {\bf d}_{n-k}\cdot{\bf d}_k \right] \, , \qquad
\bG_n = \sum_{k\in \bZ} \bfa_{k} \cdot {\bf d}_{n-k}\, , 
\ee
and similarly for $(\tilde{\bL}_n,\tilde{\bG}_n)$. 

The gauge fixed action depends only on the zero modes and the transverse oscillator variables, and is 
\begin{eqnarray}\label{RamondD}
S &=& \int \! dt \left\{\dot x^m p_m  + \frac{i}{2} \left({\bf d}_0 \cdot \dot {\bf d}_0 + \tilde{\bf d}_0 \cdot \dot{\tilde {\bf d}}_0 \right)
+ \sum_{k=1}^\infty \frac{i}{k} \left(\bfa_{-k}\cdot\dot\bfa_k + \tbfa_{-k} \cdot \dot{\tilde\bfa}_{-k} \right) \right. \nonumber \\
&& \left.  + \sum_{k=1}^\infty i\left({\bf d}_{-k}\cdot \dot {\bf d}_k + \tilde{\bf d}_{-k}\cdot \dot {\tilde{\bf d}}_k \right) 
\quad  -  \lambda_0L_0 - \tilde\lambda_0 \tilde L_0 \right\}\, , 
\end{eqnarray}
where
\be
L_0 = \frac{1}{8\pi T} \left( p^2  +  8\pi T N\right) \, , \qquad \tilde L_0 = \frac{1}{8\pi T} \left( p^2  +  8\pi T \tilde N\right)\, , 
\ee
with level numbers
\be
N= \sum_{k=1}^\infty \left( \bfa_{-k}\cdot\bfa_k + k\,{\bf d}_{-k}\cdot{\bf d}_k\right)\, , \qquad 
\tilde N = \sum_{k=1}^\infty \left( \tbfa_{-k}\cdot\tbfa_k + k\, \tilde{\bf d}_{-k}\cdot \tilde{\bf d}_k\right)\, . 
\ee
The two surviving constraints, imposed by the Lagrange multipliers $(\lambda_0,\tilde\lambda_0)$, are equivalent to 
the mass-shell condition 
\be\label{RamondMS}
p^2+{\cal M}^2=0 \, , \qquad {\cal M}^2 = 4\pi T \left(N +\tilde N\right)\, , 
\ee
and the level-matching condition $N= \tilde N$.

\subsection{The 3D Ramond string}

So far the space-time dimension has been arbitrary but now we focus on the 3D case. In this case there is a single tranverse direction, so the $(D-2)$-vector 
variables $(\bfa_k, {\tilde\bfa}_k)$  have only one-component; we shall therefore drop the boldface notation.  The action (\ref{RamondD}) now simplifies to 

\begin{eqnarray}\label{RamondD2}
S &=& \int \! dt \left\{\dot x^m p_m  + \frac{i}{2} \left(d_0  \dot d_0 + \tilde d_0 \dot{\tilde d}_0 \right)
+ \sum_{k=1}^\infty \left[\frac{i}{k} \left(\alpha_{-k}\dot\alpha_k + \tilde\alpha_{-k} \dot{\tilde\alpha}_{-k} \right) + i\left(d_{-k}\dot d_k + \tilde d_{-k} \dot {\tilde d}_k \right) \right]
\right. \nonumber \\
&&\left. 
\quad  - \  \frac{\lambda_0}{8\pi T} \left(p^2 + 8\pi T N\right)  - \frac{\tilde\lambda_0}{8\pi T} \left(p^2+ 8\pi T \tilde N\right) \right\}\, , 
\end{eqnarray}
where the level-number operators are now 
\be
N= \sum_{k=1}^\infty \left(\alpha_{-k}\alpha_k + k\, d_{-k} d_k\right)\, , \qquad \tilde N= \sum_{k=1}^\infty \left(\tilde\alpha_{-k}\tilde\alpha_k + k\, \tilde d_{-k} \tilde d_k\right)\, . 
\ee
In this simplified 3D notation the operators $(\bL_n,\bG_n)$ of (\ref{transverseL})  are
\be\label{transverseL3}
\bL_n = \frac{1}{2}\sum_{k\in\bZ} \left[ \alpha_k \alpha_{n-k} + k\, d_{n-k} d_k \right] \, , \qquad \ 
\bG_n = \sum_{k\in \bZ} \alpha_{k} d_{n-k}\, , 
\ee
and similarly for $(\tilde{\bL}_n,\tilde{\bG}_n)$. At this point it is convenient  to make the following observation: the operators 
\be
\bG_0 - \alpha_0 d_0 = \sum_{k=1}^\infty \left(\alpha_{-k} d_k + d_{-k}\alpha_k\right) \, , \qquad \tilde\bG_0 - \tilde{\alpha}_0 \tilde{d}_0 = \sum_{k=1}^\infty \left(\tilde\alpha_{-k} \tilde d_k + \tilde d_{-k}\tilde\alpha_k\right) \, , 
\ee
satisfy, in the quantum theory, 
\be
(\bG_0 - \alpha_0 d_0)^2 = N\, , \qquad (\tilde\bG_0 - \tilde{\alpha}_0 \tilde{d}_0)^2 = \tilde N\, . 
\ee
There is no operator ordering ambiguity in the expressions for $(\bG_0,\tilde\bG_0)$, so these equations fix the operator ordering ambiguity in the expressions for 
$(N,\tilde N)$, in such a way that the zero-point energies cancel. In addition, if we take into account the level-matching condition ($\tilde N=N$) then we have
\be\label{Xi}
\Xi^2 = {\cal M}^2 \, , \qquad \Xi = \sqrt{8\pi T}\,  (\bG_0 - \alpha_0 d_0)\, . 
\ee
The operator $\Xi$ therefore determines the mass spectrum. It annihilates the oscillator ground state, defined to be the state annihilated by $(\alpha_k,d_k)$ and $(\tilde\alpha_k,\tilde d_k)$ for $k>0$, so the ground states are massless and all higher-level states are massive. 

Our next task is to find  the Poincar\'e Casimir operator $\Lambda$ in terms of the canonical variables of the light-cone gauge action. Prior to gauge fixing, we find that 
$\Lambda= \Lambda_++\Lambda_-$,  
where 
\be
\Lambda_+ = \frac{1}{2} \epsilon^{m}{}_{np}\,  p_m \left(-i d_0^n d_0^p  + S^{np} \right)\, , \qquad 
\Lambda_- = \frac{1}{2} \epsilon^{m}{}_{np}\,  p_m \left(-i {\tilde d}_0^n {\tilde d}_0^p  +  \tilde S^{np} \right)\, .
\ee
We will focus on $\Lambda_+$ as the calculations for $\Lambda_-$ are identical but with tilded rather than untilded quantities.  After fixing the gauge as described above and using the identity
\be
\sum_{n\in \bZ} \bG_{-n} d_n \equiv 0\, , 
\ee
we find that\footnote{Since $\epsilon ^{012} = 1$, in our conventions, we have $\epsilon ^{+ - 2} = 1$.} 
\bea\label{LambdaRNS}
\Lambda_+ &=&  p_- \left(id_0^-d_0 -S^{- 2} \right) \nonumber \\
&=&  \sqrt{4\pi T}\sum_{n = 1}^\infty \frac{i}{n}\left(\bL_n\alpha_{-n} -\alpha_n\bL_{-n}\right)\, . 
\eea
Notice that this expression includes a term linear in $d_0$ coming from the $k=n$ term in the sum of (\ref{transverseL3}) for $\bL_n$.

We have now constructed the operators necessary to determine the spectrum of the closed 3D Ramond string. We will not construct the spectrum explicitly; instead, we will demonstrate that it is the same as that of the closed 3D GS superstring.

\section{Equivalence to the 3D GS superstring}

We now want to compare the results obtained above for the closed 3D Ramond closed string with those obtained in \cite{Mezincescu:2010yp} for the closed 3D Green-Schwarz superstring. The light-cone gauge conditions imposed in  \cite{Mezincescu:2010yp} included the condition $x^+(t)=t$, but it is simpler to allow the zero-mode $x^+(t)$ to remain an 
arbitrary function of time, at the cost of maintaining in the action the mass-shell constraint, as we did above for the Ramond string. The mass-squared operator can then be read off 
from this additional constraint. The bosonic canonical variables  in light-cone gauge are obviously the same as those of the Ramond string, so we here adopt the same notation for them.
Taking these  minor modifications into account, we may use the results of  \cite{Mezincescu:2010yp}  to write down the Fourier-space action in light-cone gauge for the closed 3D GS superstring; it is 
\begin{eqnarray}
S &=& \int dt \left\{ \dot x^m \mathcal{P}_m + \frac{i}{2} \vartheta_a \dot\vartheta_a + i \sum_{k=1}^\infty \left[ \frac{1}{k} \left(\alpha_{-k} \dot \alpha_k 
+ \tilde\alpha_{-k} \dot {\tilde\alpha}_k\right) + \xi_{-k} \dot\xi_k + \tilde\xi_{-k} \dot {\tilde \xi}_k \right]\right. \nonumber \\
&&\left. \quad - \  \frac{1}{4\pi}\ell_0\left(\mathcal{P}^2 + {\cal M}^2\right) - u_0 \left(\tilde N - N\right)\right\}\, , 
\end{eqnarray}
where the Poincar\'e charges and mass-squared operator are\footnote{The string tension parameter  $T$  of  \cite{Mezincescu:2010yp} is actually $2\pi$ times the tension, so we have rescaled the result of  \cite{Mezincescu:2010yp} to take this into account.}
\be
{\cal P}_+ = p_+ - \frac{2\pi T}{p_-} \left(\nu +{\tilde \nu}\right)\,, \qquad {\cal P}_- = p_- \,, \qquad {\cal P}_2 = p_2 \,, 
\ee
and 
\be
{\cal M}^2 = 4\pi T \left(N+\nu+{\tilde N} +{\tilde \nu}\right) \, , 
\ee
with 
\be
N= \sum_{k=1}^\infty \alpha_{-k} \alpha_k ,\quad \nu = \sum_{k = 1}^\infty k \, \xi_{-k}\xi_k  \, , \quad
\tilde N= \sum_{k=1}^\infty \tilde\alpha_{-k}\tilde\alpha_k, \quad {\tilde \nu}= \sum_{k =1}^\infty k\, \tilde\xi_{-k}\tilde\xi_k\, . 
\ee
It is also useful to recall here the operators $\Xi$ and $\tilde{\Xi}$, which satisfy $\Xi^2 = \mathcal{M}^2 = \tilde{\Xi}^2$:
\be
\Xi = \sqrt{8 \pi T}\sum_{k=1}^{\infty} (\alpha_k \xi_{-k} + \alpha_{-k} \xi_k) \,, \qquad \tilde{\Xi} = \sqrt{8 \pi T}\sum_{k=1}^{\infty} (\tilde{\alpha}_k \tilde{\xi}_{-k} + \tilde{\alpha}_{-k} \tilde{\xi}_k) \,.
\ee
If we also define\footnote{These expressions are equivalent to those given in \cite{Mezincescu:2010yp}.}
\bea
\beta_n &=& \frac{1}{2}\sum_{k\ne 0,n} \alpha_k\alpha_{n-k} \, , \qquad \gamma_n = \frac{1}{2}\sum_{k\ne0,n}\left(n-k\right)d_k d_{n-k} \, , \\
\tilde\beta_n &=& \frac{1}{2}\sum_{k\ne 0,n} \tilde\alpha_k\tilde\alpha_{n-k} \, , \qquad \tilde\gamma_n = \frac{1}{2}\sum_{k\ne0,n}\left(n-k\right)\tilde d_k \tilde d_{n-k} \, , 
\eea
then the Casimir $\Lambda$ can be written as $\Lambda= \Lambda_+ +\Lambda_-$, where  \cite{Mezincescu:2010yp} 
\bea
\Lambda_+ &=& 2\sqrt{\pi T}\sum_{k=1}^{\infty} \frac{i}{k}[ \alpha_{-k}(\beta_k + \gamma_k) - (\beta_{-k} + \gamma_{-k})\alpha_k] + \frac{i}{\sqrt{8}} \vartheta_1 \Xi \,,\nonumber \\
\Lambda_- &=& 2\sqrt{\pi T}\sum_{k=1}^{\infty} \frac{i}{k}[ \tilde{\alpha}_{-k}(\tilde{\beta}_k + \tilde{\gamma}_k) - (\tilde{\beta}_{-k} + \tilde{\gamma}_{-k})\tilde{\alpha}_k] + \frac{i}{\sqrt{8}} \vartheta_2 \Xi \,.
\eea

Now relabel the variables
\begin{gather}
x^m \mapsto x^m \,, \qquad \mathcal{P}_m \mapsto p_m \,, \qquad \xi_k \mapsto d_k \,, \qquad \tilde{\xi}_k \mapsto \tilde{d}_k \nonumber \\
\vartheta_1 \mapsto d_0 \,, \qquad \vartheta_2 \mapsto \tilde{d}_0 \,, \qquad \ell_0 \mapsto {\frac{\lambda_0 + \tilde{\lambda}_0}{2T}} \,, 
\qquad u_0 \mapsto { \frac{{\tilde \lambda}_0 - {\lambda}_0}{2} }\,.
\end{gather}
This renaming makes the lightcone-gauge GS superstring action identical to that of the Ramond string. The Poincar\'e charges, the mass-squared operator and the operators $\Xi$ and $\tilde{\Xi}$ are also identical. The only further thing required for full  quantum-mechanical equivalence is  equivalence of the Casimir $\Lambda$. To see this equivalence,  we observe 
that
\be
\beta_k + \gamma_k + \frac{k}{2}d_k =  \bL_k - \alpha_0\alpha_k \,.
\ee
and similarly for the tilded quantities. Therefore,
\bea
\Lambda_+ &=& \sqrt{4\pi T}\sum_{k=1}^{\infty} \frac{i}{k}[ \alpha_{-k}(\beta_k + \gamma_k + \frac{k}{2}d_k) - (\beta_{-k} 
+ \gamma_{-k} -\frac{k}{2}d_{-k})\alpha_k] \nonumber \\
&=& \sqrt{4\pi T}\sum_{k=1}^{\infty} \frac{i}{k}[ \alpha_{-k} (\bL_k - \alpha_0\alpha_k) - (\bL_{-k} - \alpha_0\alpha_{-k})\alpha_k] \nonumber \\
&=& \sqrt{4\pi T}\sum_{k=1}^{\infty} \frac{i}{k}(\bL_k \alpha_{-k} - \alpha_k\bL_{-k}) \,,
\eea
which is identical to (\ref{LambdaRNS}). The identity of the two $\Lambda_-$ operators, and therefore the full $\Lambda$ operators, can be similarly established. Since both 3D Poincar\'e Casimir operators are the same for the two string theories, we conclude that  the Ramond superstring and the GS superstring are quantum mechanically equivalent.

\section{Discussion}

In this paper we have obtained a  gauge-fixed form  of the closed Ramond string, in Hamiltonian form, in a general spacetime dimension $D$. In principle, the spectrum of the quantum theory can be read off from the surviving mass-shell and level-matching constraints, which become the physical-state conditions, but  Lorentz invariance is anomalous for $D\ge4$ unless $D=10$.  However, the Lorentz anomaly is also absent, trivially, for $D=3$, and in this case the spectrum is determined by the joint eigenvalues of  the operators representing the two Casimirs of the 3D Poincar\'e group, one of which gives the masses and the other the ``relativistic helicities''. What we have shown here is that, for a particular choice of canonical variables, these two operators coincide with those found previously from an analysis of the 3D closed Green-Schwarz superstring. More precisely, the closed Ramond string with local $(1,1)$ worldsheet supersymmetry is equivalent, in light-cone gauge, to the closed GS superstring with ${\cal N}=2$ spacetime supersymmetry. 

The  equivalence of the two 3D Poincar\'e Casimir operators  holds separately for the left and right Hilbert spaces of the closed strings, so our results  also imply the equivalence of the closed 3D Ramond string with only $(1,0)$ local worldsheet supersymmetry (i.e. a 3D heterotic string) to the GS superstring with ${\cal N}=1$ space-time supersymmetry. Additionally, they  imply the equivalence of GS and Ramond 3D strings with free-end boundary conditions; in this case the Ramond string has a ``hidden'' ${\cal N}=1$ space-time supersymmetry. We presume that there is some equivalence for other choices of open-string boundary conditions but this remains to be investigated.

In the critical dimension, the equivalence of the RNS and GS strings, after GSO projection,  is made possible by the ``triality'' property of representations of the transverse
${\rm Spin}(8)$ rotation group. The analogous equivalence established here of  the 3D Ramond string to the 3D GS superstring is made possible by the ``triviality'' of the transverse ${\rm Spin}(1)$ rotation group. Because all oscillator modes are singlets it is not surprising that the Ramond and GS actions can be mapped into each other. However, it is not obvious that the helicity operators of these two theories should coincide, because the way that Lorentz invariance is realized in the two theories prior to gauge fixing is quite different. Nevertheless, they do coincide. A corollary  of this coincidence is that the spectrum of the 3D closed Ramond string has particles of irrational spin  at level $3$ (and above) because this is what was found in  \cite{Mezincescu:2010yp} for the closed ${\cal N}=2$ GS superstring. 

Another  corollary is that the 3D Ramond string actually has a ``hidden'' spacetime supersymmetry, since this is an explicit feature of the 3D GS superstring, discussed in detail in  \cite{Mezincescu:2010yp}. We may use the equivalence to the GS superstring to construct the space-time supersymmetry generators of the Ramond string, at least in light-cone gauge. These are
\be
{\cal Q}_1 = \sqrt{\frac{1}{\sqrt{2} p_-}} \left(\begin{array}{c} \sqrt{4\pi T}\, \, \bG_0 \\ \sqrt{2} \, p_-\,  d_0 \end{array}\right)\, , \qquad 
{\cal Q}_2 = \sqrt{\frac{1}{\sqrt{2} p_-}} \left(\begin{array}{c} \sqrt{4\pi T}\, \, \tilde{\bG}_0 \\ \sqrt{2} \, p_-\, \tilde d_0 \end{array}\right)\, .
\ee
It may be checked, using the  (anti)commutation relations of  canonical variables of the light-cone gauge-fixed Ramond string, that these supersymmetry charges 
satisfy the 3D ${\cal N}=2$ supersymmetry algebra; the conventions that are needed for this check are  detailed  in \cite{Mezincescu:2010yp}.

The Lorentz anomaly of a generic subcritical-dimension string quantized in light-cone gauge is usually considered to be another manifestation of the conformal anomaly 
that arises from quantization in conformal gauge, where Lorentz invariance is manifest.  Conformal  invariance in conformal gauge is a residual gauge invariance, implying a residual redundancy, and the  cost for removing this redundancy in a subcritical dimension $D>3$ is a Lorentz anomaly. The $D=3$ situation is quite different, however, because there is no Lorentz anomaly,  at least not if the Lorentz group $SO(1,2)$ is replaced by its universal cover.  This suggests that there should be an alternative  Lorentz covariant quantization prescription,  available only for $D=3$,  such that the usual conformal anomaly is absent.  

Another, possibly related, issue is modular invariance of the one-loop partition function. As mentioned in the introduction, the fact that we need only the RR sector of the 
3D RNS string for equivalence with the 3D GS superstring contrasts with the fact that modular invariance of the critical-dimension RNS string requires the inclusion of  all sectors.  To address this  we must first discuss  the logically prior issue of  modular invariance of the {\it bosonic} 3D string: the standard computation shows that modular-invariance holds only in the critical dimension but  $D=3$ is again special.  The standard computation starts from  a flat Euclidean torus embedded in a $D$-dimensional Euclidean space but a flat torus cannot be smoothly embedded into this space when $D=3$. This is perhaps analogous to the fact that closed worldlines can be knotted when $D=3$, which is essentially how the possibility of anyons arises; and anyons are indeed present in the spectrum of any 3D string. 

Given that the 3D bosonic string could be modular invariant because of some additional contributions to the path integral arising from the above observation, we could 
then ask how the RR sector of the  3D RNS string could be modular invariant  by itself. Here we take space-time supersymmetry as our guide.  When space-time supersymmetry is manifest the worldsheet `fermions'  must be periodic;  in particular, the GS fermions are always periodic. In the critical dimension,  the RNS superstring is not manifestly supersymmetric, even in light-cone gauge, and all sectors  are needed both for  equivalence to the GS string and for modular invariance. In the 3D case, we actually have something stronger than the critical dimensional ``equivalence''  because the RR sector of the 3D RNS string  in light-cone gauge ``is''  the 3D GS superstring in light-cone gauge. From this perspective, it is obvious that we must keep only the Ramond sector of the RNS string because that is what we must do for the GS superstring.  

Having made this point, we should perhaps conclude with the following caveat. It was pointed out in \cite{Bengtsson:1987hq} that there is a potential problem with the Green-Schwarz action because first-quantization of any relativistic theory leads to both positive and negative energy states whereas manifest spacetime supersymmetry implies positive energy, assuming the absence of negative norm states. The same point 
applies equally to the superparticle, of course,  and the resolution there, pointed out in \cite{Gauntlett:1990xq} in the context of a particular unitary gauge, is that there is a correlation (after gauge-fixing) between the sign of the energy and the sign of the fermion kinetic terms, and this correlation ensures that the standard supersymmetry algebra holds for both the positive energy and the negative energy states. In the light-cone gauge fixing of the 3D GS superstring detailed in \cite{Mezincescu:2010yp}  it was assumed that $p_-$ is positive because this makes the energy positive, but it follows from the results found there that the sign of $p_-$ is correlated with the sign of the fermion kinetic terms. This is as expected, for the reasons just given, but there is no analogous sign correlation for the Ramond string. It is not clear to us what the implications of this difference are, but we remark that the same observation applies to the RNS/GS equivalence in  the critical dimension.

\section*{Acknowledgements} L.M. acknowledges partial support from National Science Foundation Awards PHY-0855386 and PHY-1214521. 
A.J.R. acknowledges support from the UK Science and Technology Facilities Council. We are grateful to Michael Green and Dieter L\"ust for helpful conversations and correspondence.


\providecommand{\href}[2]{#2}\begingroup\raggedright\endgroup

\end{document}